\documentclass[12pt,preprint]{aastex}

\begin{document}

\title{Unraveling the nature of coherent pulsar radio emission}
\author{Dipanjan Mitra}
\affil{National Centre for Radio Astrophysics Ganeshkhind, Pune 411 007 India}
\author{ Janusz Gil and George I. Melikidze\altaffilmark{1}}
\affil{J.Kepler Institute of Astronomy, University of Zielona Gora, Poland}
\email{dmitra@ncra.tifr.res.in,jag@astro.ia.uz.zgora.pl,gogi@astro.ia.uz.zgora.pl} \altaffiltext{1}{Georgian National
Astrophysical Observatory, Chavchavadze State University, Kazbegi 2a, Tbilisi, Georgia}

\begin{abstract}
Forty years have passed since the discovery of pulsars, yet the
physical mechanism of their coherent radio emission is a
mystery. Recent observational and theoretical studies strongly suggest
that the radiation outcoming from the pulsar magnetosphere consists
mainly of extraordinary waves polarized perpendicular to the planes of
pulsar dipolar magnetic field. However, the fundamental question
whether these waves are excited by maser or coherent curvature
radiation, remains open. High quality single pulse polarimetry is
required to distinguish between these two possible mechanisms. Here we
showcase such {\it decisive} strong single pulses from 10 pulsars
observed with the GMRT, showing extremely high linear polarization
with the position angle following locally the mean position angle
traverse. These pulses, which are relatively free from depolarization,
must consist of exclusively single polarization mode. We associate
this mode with the extraordinary wave excited by the coherent
curvature radiation. This crucial observational signature enables us
to argue, for the first time, in favor of the coherent curvature
emission mechanism, excluding the maser mechanism.
\end{abstract}


\section{Introduction}

Pulsar coherent radio emission originates within the flux tube of the
open force lines of dipolar magnetic field. This conclusion is
unavoidable from the theoretical point of view as these are the only
field lines that can develop an ultra-high potential drop acting as a
basic source of pulsar activity (Goldreich \& Julian 1969). As the
pulsar emission beam sweeps past the observer a single radio pulse is
observed, which typically consists of one to several subpulses. Each
subpulse is emitted within a sub-bundle of the overall flux tube of
the open field lines. When a large number of single pulses are added
together in phase, then a stable mean or average profile is
obtained. These profiles consist of one to several components, with an
actual number depending on how close the observer's line-of-sight
(LOS) approaches the pulsar beam axis. In the case of small approach
angle (called the impact angle), the mean profile consists of a
central (core) component, usually flanked by one or two pairs of conal
components (Backer 1976, Rankin 1993). The position angle (PA) of the
linear polarization across the average pulse profile shows a
characteristic S-swing or traverse, which is associated with a range
of open magnetic field line planes intersected by LOS (Radhakrishnan
\& Cooke 1969). This swing is usually steep under the core component,
but is much shallower or even flat under the outer conal
components. The model describing this swing is commonly known as the
rotating vector model (RVM).  Canonically, the RVM holds for average
profiles, without specifying what kind of PA variation should be
expected for a single pulse (subpulse) emission. We address this
problem below in this paper.

Early polarimetric studies revealed that single pulses in pulsar
radiation are highly linearly polarized, with moderate, sign-changing
circular polarization observed in some cases (see for e.g. Clark \&
Smith 1969, Lyne, Smith \& Graham 1971, Manchester, Taylor \& Huguenin
1975). However, none of these observations were transparent enough to
pin down the emission mechanism. Here we present a set of high quality
single pulse polarimetric observations from the Giant Meterwave Radio
Telescope (GMRT, Swarup et al. 1991) with an aim to identify the
pulsar emission process.

\section{Results}

A bright subpulse in a single pulse from the radio pulsar PSR B1237+25
observed at 325 MHz is shown in Fig.~1. The subpulse, which appears at
the conal region, is close to a gaussian shape, with a full width half
maximum of about 0.7$^{\circ}$. It is very highly linearly polarized
(93\% at the peak) and the circular polarization changes sign close to
the subpulse maximum. The most notable finding is that the {\it PA of
  linear polarization in subpulses follow closely the mean PA curve at
  the corresponding profile components.} Generally, when a highly
polarized single pulses appear the subpulse PA variations follow the
mean PA traverse. We use this crucial observational feature in our
data to identify the mechanism of pulsar radio emission. Fig.~2 shows
similar subpulses from 9 more pulsars, all of them are the conal
components of their respective pulses.

\begin{figure}[ht]
\includegraphics[width=54mm,angle=-90.]{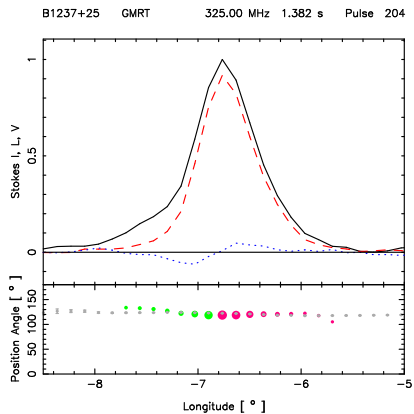}
\figcaption{The plot shows a strong subpulse in a single pulse of PSR
  B1237+25. The top panel of the plot shows a symmetric pulse which is
  highly linearly polarized, has a sign changing circular where the
  peak of the total intensity (in black), peak of linear (red)
  polarization and zero of circular (blue) lie close to each
  other. The bottom panel shows the position angle (PA) traverse for
  the average PA traverse obtained from a larger set of pulses in
  light gray and the PA traverse of the subpulse is shown in green and
  magenta colors corresponding to negative and positive circular
  polarization respectively. This signature is typical of curvature
  radiation as discussed in the text.}
\label{fig1}
\end{figure}

The observations were done at 325 MHz with the phased array mode of
the GMRT. A bandwidth of 16 MHz was used and the data was recorded at
a sampling interval of 0.512 ms. About 2\% accuracy in the stokes
parameters were obtained by applying the polarization calibration
procedure (Mitra, Gupta \& Kudale 2005). The calibrated stokes were
used to construct the linear polarization L =
$\sqrt{\rm{U}^2+\rm{Q}^2}$ and the PA = 0.5
tan$^{-1}({\rm{U}}/{\rm{Q}})$. The convention followed for the
circular polarization V is Left--Hand--Circular - Right--Hand--Circular.

\section{Discussion}

Before we proceed to understand the implications of the observations
in Figs.~1 and 2, we need to discuss how coherent radio emission can
originate and escape from the magnetospheric plasma. Generally
speaking, the coherent pulsar radio emission should be generated by
means of either a maser or coherent curvature mechanism (Ginzburg \&
Zheleznyakov 1975, Kazbegi, Machabeli \& Melikidze 1991). This
radiation, while propagating in the magnetosphere splits naturally
into the ordinary and extraordinary waves, which correspond to the
normal modes of strongly magnetized plasma (see e.g.  Arons \& Barnard
1986). The ordinary waves are polarized in the plane of the wave
vector $k$ and the local magnetic field direction and their electric
field has a component along the magnetic line of force. Therefore they
interact strongly with plasma particles and thus encounter difficulty
in escaping the magnetosphere. On the other hand, the extraordinary
waves are linearly polarized perpendicularly to the wave vector $k$
and the local magnetic field. As a result they can propagate through
the magnetospheric plasma almost as in vacuum and thus reach the
observer (see Gil, Lyubarsky \& Melikidze 2004, (hereafter GLM04) for
a detailed discussion on the nature of ordinary and extraordinary
waves in pulsar magnetosphere).

\begin{figure*}[t]
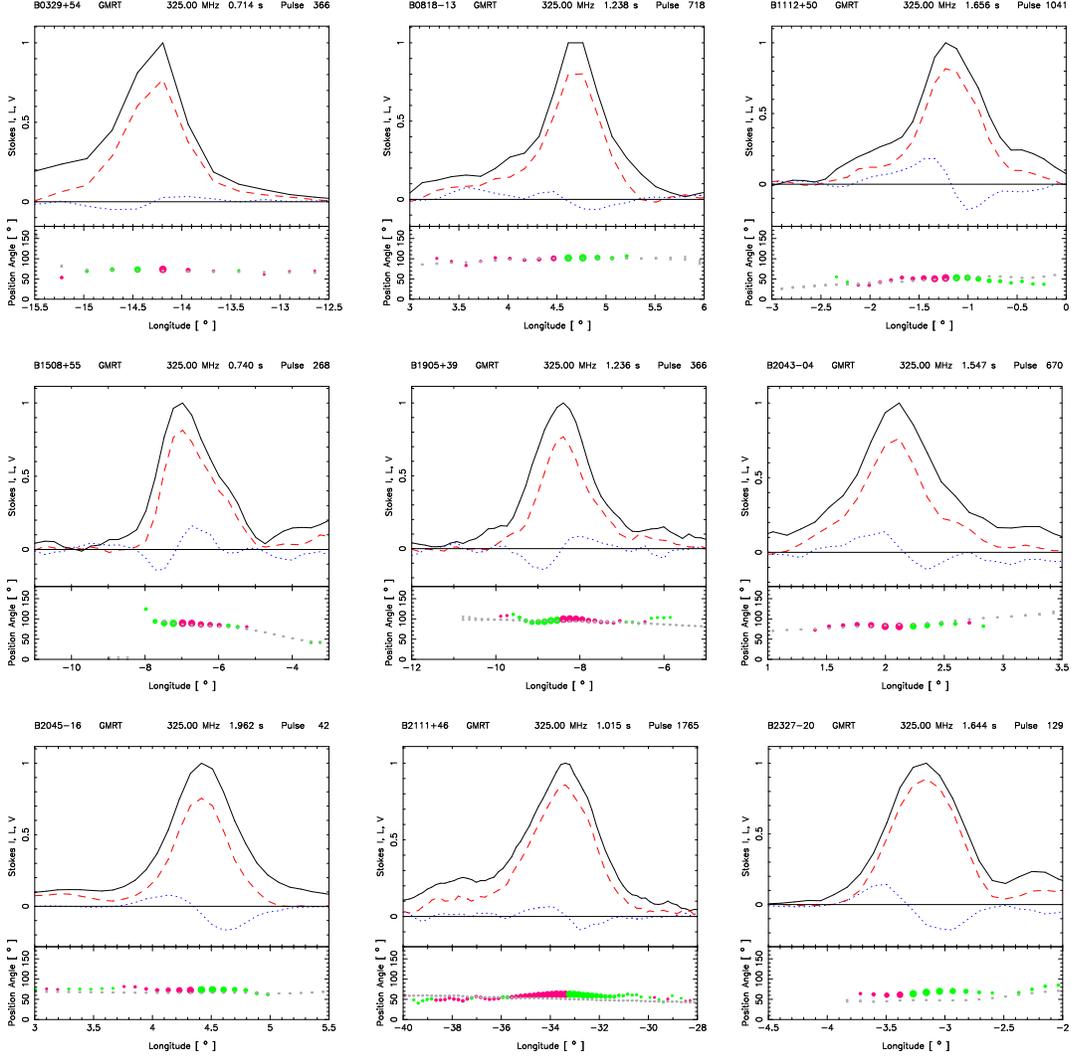

\begin{tabular}{@{}lrr@{}}
{\mbox{\includegraphics[width=44mm,angle=-90.]{PQB0329+54_SP1.ps}}}&
{\mbox{\includegraphics[width=44mm,angle=-90.]{PQB0818-13_SP1.ps}}}&
{\mbox{\includegraphics[width=44mm,angle=-90.]{PQB1112+50_SP1.ps}}}\\
{\mbox{\includegraphics[width=44mm,angle=-90.]{PQB1508+55_SP2.ps}}}&
{\mbox{\includegraphics[width=44mm,angle=-90.]{PQB1905+39_SP1.ps}}}&
{\mbox{\includegraphics[width=44mm,angle=-90.]{PQB2043-04_SP1.ps}}}\\
{\mbox{\includegraphics[width=44mm,angle=-90.]{PQB2045-16_SP1.ps}}}&
{\mbox{\includegraphics[width=44mm,angle=-90.]{PQB2111+46_SP1.ps}}}&
{\mbox{\includegraphics[width=44mm,angle=-90.]{PQB2327-20_SP1.ps}}}\\
\end{tabular}
\caption{Same as in Fig.~1. The subpulses are from pulsars PSRs B0329+54, B0818-13, B1112+50, B1508+55, B1905+39,
B2043-04, B2045-16, B2111+46 \& B2327-20 (the order is from left to right and top to bottom).}
\label{fig2}
\end{figure*}

  Interestingly, the X-ray image of the Vela pulsar can give an
   insight to the pulsar's emission geometry.  The absolute orientation
   of the polarization plane found in this case definitely
   demonstrates that the polarization direction of radio waves from
   the Vela pulsar is perpendicular to the planes of dipolar magnetic
   field lines (Lai, Chernoff \& Cordes 1991). Therefore, undoubtedly
   this radiation consists of extraordinary waves (see also discussion
   in Section 6.4 in GLM04). Also, based on the assumption that
   pulsar's proper motions are parallel to the rotation axis (Johnston
   et al. 2005, Rankin 2007), it was argued that the primary
   polarization mode for pulsar PSR B0329+54 corresponds to the
   extraordinary mode (Mitra, Rankin \& Gupta 2007). These conclusions
   strengthen the argument that the observed pulsar radiation consists
   mainly of extraordinary waves polarized perpendicular to the planes
   of dipolar magnetic field lines.  It is worth emphasizing that for
   the first time the radiation of point like charge moving along
   the curved magnetic field lines in the relativistic
   electron-positron plasma of the pulsar magnetosphere was
   self-consistently treated by GLM04. They found the exact solution
   of the corresponding set of Maxwell equations in the far zone
   to be the extraordinary wave mode. Thus, they demonstrated, that the
   extraordinary mode can be generated in strong curved magnetic
   field via the linear coupling of the normal modes in the radiation
   formation region (see also Section 5 in GLM04).

However, an open question remains i.e. what kind of input coherent
radiation excites these waves emanating from the pulsar?  This
question can be answered based on the highly polarized subpulse for 10
pulsars presented in this paper.  Theoretically, excitation of
escaping waves in pulsar magnetosphere is possible either by maser or
coherent curvature emission mechanisms. The subpulse width in the case
of maser mechanism corresponds to the opening angle of the maser
emission which depends on the resonant conditions necessary for the
plasma instability to be developed (Kazbegi, Machabeli, Melikidze \&
Smirnova 1991). In maser radiation the $k$ vector can be oriented in
any direction with respect to the local magnetic field. As we already
mentioned, the electric vector of the extraordinary waves is
perpendicular to the plane containing both $k$ and magnetic field
vector $B_0$, while the position angle of the ordinary waves lies in
this plane. Thus, in case of maser radiation, the PA (orientation of
electric field vector $E$) across the subpulse width should perform
swings, rather than remain tightly aligned with the mean PA traverse
which reflects the orientation of the magnetic field plane, as per
RVM. This is illustrated for the extraordinary mode in the upper panel
of Fig.~3, where the electric field $E$ changes direction for
different viewing angles defined by $k$ vectors. In contrast, the
curvature radiation is highly collimated along the local magnetic
field $B_0$. Additionally, the electric vector excited by the coherent
curvature emission in plasma must be either perpendicular or parallel
to the plane of curvature of the magnetic field lines. Thus, the
subpulse PA variation will reflect the change in the orientation of
the magnetic field planes, that is the subpulse PA will closely follow
the RVM-like mean PA traverse. This is illustrated in the lower panel
of Fig.~3 for extraordinary waves, where the electric vectors $E$ are
perpendicular to the planes of magnetic field lines. Our observations
clearly demonstrate that the observed PA variation across the subpulse
follow the mean PA (RVM-like) traverse, implying that the observed
emission is due to waves excited by the coherent curvature
radiation. As already mentioned, from a theoretical perspective, only
the extraordinary wave can escape the pulsar magnetosphere freely, and
hence we conjecture that these strong highly polarized single pulses
are likely to be freely escaping extraordinary waves excited by
coherent curvature radiation.

High linear polarization and sense-reversing circular polarization is
a property of curvature radiation from a single charged particle
moving relativistically along curved magnetic field lines (Michel
1987). This idea had led to the development of the theory of curvature
radiation (Gil \& Snakowski 1990 a,b) of small hypothetical charged
bunch emitting coherently (Ruderman \& Sutherland 1975). In fact, for
a gaussian intensity envelope this model can faithfully reproduce the
observations reported in Fig.~1 and 2 (Gil, Kijak \& Zycki 1993). This
theory, however, had serious problems. First, it was developed for
vacuum, without considering generation and propagation of the emitted
radiation in pulsar magnetospheric plasma. Second, the mechanism for
formation of elementary bunches that could emit coherent curvature
radiation was unknown. The general opinion was that formation of such
bunches was difficult (see Melrose 1995 for review).

\begin{figure}[ht]
\includegraphics[width=80mm,angle=90.]{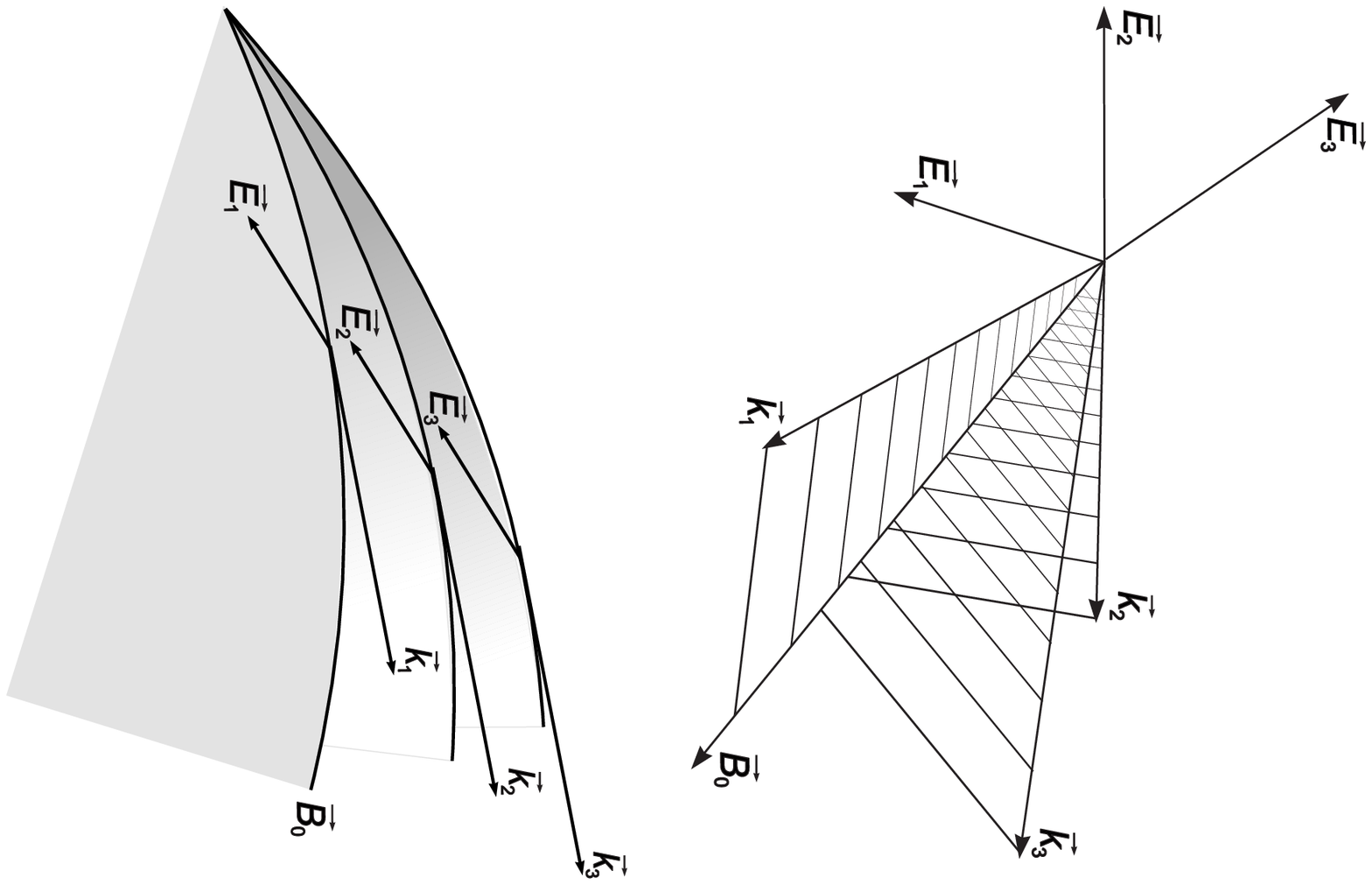}
\caption{Behavior of the wave electric field vector (i.e. the position
  angle) in the case of maser (upper panel) and curvature (lower
  panel) emission mechanisms. In the upper panel the position angle
  alters fast across the subpulse width, while in the lower panel the
  PA variation are determined by the range of orientations of the
  planes of dipolar magnetic field lines encompassed by a beam of
  subpulse emission.} \label{fig3}
\end{figure}

Let us now briefly describe the theoretical model capable of forming
elementary bunches emitting the coherent curvature radiation. Close to
the neutron star where the observed radio emission is found to
originate ( which is about 300-500 km, see for e g. Rankin 1993,
Blaskiewics, Cordes \& Wasserman 1981, Kijak \& Gil 1997), the well
known two-stream plasma instability naturally generates the Langmuir
plasma waves (e.g. Asseo \& Melikidze 1998). Consequently, a
spark-associated soliton model of the coherent pulsar radio emission
has been developed (Melikidze, Gil \& Pataraya 2000), in which the
non-stationary sparking discharge of the ultra-high potential drop
just above the polar cap results in modulational instability of
Langmuir waves. This leads to formation of small, relativistic,
charged solitons, able to emit coherent curvature radiation. So, a
natural mechanism for the formation of charged bunches was found. The
only deficiency of the soliton model was that the influence of the
ambient plasma on the formation and propagation of the emitted
radiation was not considered. Therefore, it was not known whether this
radiation can emerge from the pulsar and reach the observer.

This problem has now been addressed and GLM04 found that the power of
coherent curvature radiation of a point-like charge (a model of
charged soliton) moving relativistically along curved magnetic field
lines through the pulsar magnetospheric electron-positron plasma is
largely suppressed compared with the vacuum case (in which the power
was too high compared with observations). However this power is still
at a considerable level to explain the observed pulsar
luminosities. The outgoing waves are polarized perpendicularly to the
plane of curvature of magnetic field lines, and they represent the
escaping extraordinary waves. The polarization properties of the
subpulses from 10 pulsars presented in this paper strongly support
this theory. Naturally, the above statement is true provided that
propagation effects in magnetospheric plasma cannot change the
polarization state. As was shown by Cheng \& Ruderman (1979) the
subpulse polarization patterns can be, in general, affected by the
propagation effects if the so called adiabatic walking condition is
satisfied. However, a more rigorous treatment for the radiation
mechanism considered here demonstrates that the adiabatic walking
condition is not satisfied (see eq. [31] and the corresponding
discussion below it in GLM04). Therefore the waves escape from the
plasma retaining the initial polarization in the direction
perpendicular to the magnetic field line planes, exactly as it is
observed in strong and highly linearly polarized subpulses presented
in our Fig. 2.

The characteristic Lorentz factors of emitting bunches/solitons should
be about $\gamma\simeq 400$, for obtaining the observed frequency and
power due to coherent curvature radiation in the magnetospheric plasma
(GLM04).  The typical subpulse widths in Fig.~1 and ~2 is about $
1^\circ \simeq 2\times 10^{-2}$ radians, which is several times larger
than the width of the radiation cone of an elementary curvature
emitter $1/400=0.0025$ radians.  Thus the subpulse should be formed by
the incoherent sum of radiation emitted by a number of solitons
filling the flux tube of dipolar field lines with an angular extent of
about 0.02 radians in the emission region (which for a typical pulsar
with a period of 1 sec originates at an emission altitude of about 50
stellar radii, see for e.g. Melikidze et al. 2000, Kijak \& Gil 1997).
This angular width projected onto the polar cap surface gives about 1
percent of the fractional area, which is consistent with the model in
which the base of the subpulse flux tube is formed by sparks of
electron-positron avalanches (e.g. Ruderman \& Sutherland 1975, Gil \&
Sendyk 2000). This leads to generation of coherent curvature radiation
as proposed in the spark associated soliton model by Melikidze, Gil \&
Pataraya (2000).

In this paper we analyse a selection of high quality, almost
completely polarized single pulses from a number of pulsars. We argue
that in those cases we observe almost exclusively one polarization
mode, which we associate with the extraordinary waves excited by the
coherent curvature radiation (GLM04). It should be mentioned that some
earlier observations have already found evidence that single pulse PA
variations follow the mean PA traverse (see Ramachandran et
al. 2002). However those observations do not reveal any highly
polarized pulses as shown in this paper. Single pulse depolarization
can result from incoherent addition of emission overlapping from
adjacent field lines w.r.t the LOS, presence of orthogonal modes and
also propagation effects.  Our almost completely polarized pulses, are
relatively free of depolarization and hence can be associated with one
of the polarization mode. In this sense they seem to be ideal to
unravel the nature of the pulsar radio emission process. Naturally,
based on our selected data we are not able to examine the phenomenon
of orthogonally polarized modes. Hence, the question about the origin
of orthogonal modes observed in pulsar radio emission is still
open. While the answer is not yet clear, we speculate that the usually
weaker secondary orthogonal polarization mode is somehow connected
with the other (ordinary) mode excited via the coherent curvature
radiation. However, there is a theoretical problem to be solved as how
the waves polarized in the plane of the curved magnetic field can
escape from the magnetosphere?

Finally, a comment on the apparent circular polarization, reversing
sense at or near the subpulse maximum, as evident from Figs. 1 and
2. Indeed, the emitted extraordinary waves are purely linearly
polarized and the question is whether this is always true for the
observed radiation. In case of pulsars the planes of the source motion
along curved field lines rotate with respect to the observer and we
claim that the observed radiation attains the net circular
polarization for geometrical reasons, as discussed in Gil \& Snakowski
(1990 a,b). Indeed, sense reversing circular polarization in vacuum
results from the fact that each source of coherent curvature radiation
is viewed from both sides of the plane of their motion as the
observer's LOS cuts through a cones of emission (Gil, Kijak \& Zycki
1993).

A proper understanding of the pulsar emission mechanism critically
depends on our ability to analyze highly turbulent non-linear behavior
of plasma. The single pulse polarization in pulsars in conjunction
with the spark associated soliton model seems to explain one of the
most intriguing phenomenon in astrophysics. We claim that the observed
pulsar signals consists mainly of extraordinary waves (at least in the
case of strong linearly polarized subpulses) excited in magnetospheric
plasma by coherent curvature radiation. It is likely that this
development presented here will find application in several other
coherent emission process in astrophysics, like giant pulses, RRATs
and extrasolar radio bursts.

\acknowledgments We thank an anonymous referee for constructive
criticism that helped to improve our paper. We also thank Gopal
Krishna, Dipankar Bhattacharya \& Joanna Rankin for their helpful
comments on the manuscript. JG and GM acknowledges a partial support
of Polish Grants N N 203 2738 33 and N N 203 3919 34. GM was partially
supported by the Georgian NSF ST06/4-096 grant. The GMRT is run by
National Centre of Radio Astrophysics of the Tata Institute of
Fundamental Research.

\end{document}